\documentclass[pdftex,twocolumn,epjc3]{svjour3}          

\RequirePackage[T1]{fontenc}

\smartqed  

\RequirePackage{graphicx}
\RequirePackage{mathptmx}      
\RequirePackage{flushend}
\RequirePackage[numbers,sort&compress]{natbib}
\RequirePackage[colorlinks,citecolor=blue,urlcolor=blue,linkcolor=blue]{hyperref}
\RequirePackage{xcolor}
\RequirePackage[normalem]{ulem}
\RequirePackage{amsmath}

\journalname{Eur. Phys. J. A}

\begin{document}

\title{Beyond the relaxation time approximation}

\author{Grzegorz Wilk\thanksref{e1,addr1}
        \and
        Zbigniew W\l odarczyk\thanksref{e2,addr2}
}

\thankstext{e1}{e-mail: grzegorz.wilk@ncbj.gov.pl}
\thankstext{e2}{e-mail: zbigniew.wlodarczyk@ujk.edu.pl}

\institute{National Centre for Nuclear Research, Department of Fundamental Research, 02-093 Warsaw,
           Poland\label{addr1}
          \and
          Institute of Physics, Jan Kochanowski University, 25-406 Kielce, Poland\label{addr2}
}

\date{Received: date / Accepted: date}

\maketitle

\begin{abstract}
The relaxation time approximation (RTA) is a well known method of describing the time evolution of a statistical ensemble by linking distributions of the variables of interest at different stages of their temporal evolution.
We show that if all the distributions occurring in the RTA have the same functional form of a quasi-power Tsallis distribution the time evolution of which depends on the time evolution of its control parameter, nonextensivity $q(t)$, then it is more convenient to consider only the time evolution of this control parameter.
\end{abstract}

\section{Introduction}
\label{sect.intro}

Many problems in science involve understanding the time evolution of a statistical ensemble. Here we will focus on a system of particles described by the probability distribution $f(r,p,t)$ which depends on position $r$, momentum $p$ and time $t$. In general, for an evolving physical system operating irreversibly out of thermodynamical equilibrium $f(r,p,t)$ will differ from that of a Boltzmannian ensemble and its evolution is usually studied using the Boltzmann transport equation (BTE), the general form of which is \cite{Huang},
\begin{equation}
\frac{df(r,p,t)}{dt} = \frac{\partial f}{\partial t} + \vec{u}\cdot\nabla_r f + \vec{F}\cdot \nabla_p f = C[f], \label{BE}
\end{equation}
where $\vec{F}$ is the external force, $\vec{u}$ the velocity and $C[f]$ the collision term. Narrowing our interest to situations where the considered system is homogeneous (i.e. $\nabla_r f=0$) and when no external forces are acting (i.e. when $\vec{F}=0$) Eq. (\ref{BE}) simplifies to the form
\begin{equation}
\frac{df(r,p,t)}{dt} = \frac{\partial f}{\partial t} = C[f(t)]. \label{RBE}
\end{equation}
Because of the freedom in choosing the functional form of $C[f(t)]$ it is still a very general equation that allows one to deal with a variety of situations. However, usually in many applications it is further simplified to a form called the relaxation time approximation (RTA) which consists in using such a simple form of the collision term  \cite{Huang,BGK,AW,FR}:
\begin{equation}
C[f] = \frac{f_{eq} - f}{\tau}, \label{RTA}
\end{equation}
where $f_{eq}$ is the local equilibrium distribution and $\tau$ is the relaxation time, understood as the time taken by the non-equilibrium system to reach equilibrium. In this approximation the BTE simplifies further to
\begin{equation}
\frac{\partial f}{\partial t} = \frac{f_{eq} - f}{\tau}. \label{SRBE}
\end{equation}
Solving this equation for the initial conditions such that at $t=0$ one has as initial (assumed) distribution, $f = f_{in}$, and at freeze-out time, $t=t_f$ one has a final distribution, $f=f_{fo}$ (which we identify with the distribution we are looking for that actually describes the distribution obtained experimentally)\footnote{The statistical system produced in multiparticle production processes quickly reaches an initial distribution (pre-equilibrium state) which slowly evolves to equilibrium but becomes frozen at the freeze-out time (usually the system at freeze-out is not in thermodynamic equilibrium). The experimentally measured spectra of the produced particles reflect the state of the system at freeze-out.} one finds that
\begin{equation}
f_{fo} = f_{eq} + \left(f_{in} - f_{eq}\right) \exp\left( - \frac{t_f}{\tau}\right). \label{RTA-sol}
\end{equation}
The continued popularity of such an approach to the analysis of various particle production processes can be proved by the fact that recently the Boltzmann transport equation in the RTA approximation was used to analyze various observables in nucleus-nucleus collisions (in particular, to study the time evolution of temperature fluctuations in a non-equilibrium system \cite{BGSS}, to describe the elliptic flow \cite{YTTS}, transverse momentum spectra \cite{ZLD}, as well as for the study of nuclear modification factors at RHIC and LHC energies \cite{TKTS, TBGKSC, QCGZZ}).

\section{The two-component nature of the RTA approximation}
\label{sect.two-component}

In what follows we consider (for simplicity) massless objects in a one-dimensional statistical ensemble and assume that all probability distributions have the form $f(z,t)$, where $z$ is a scaled variable, $z=x/x_0$, for observable $x$ with $x_0$ being the corresponding scale parameter. The evolution of our system reaches asymptotically, for $t_f \to \infty$,  a state of local equilibrium in which $f_{fo} = f_{eq}$. In such a case
\begin{equation}
f_{eq} \left( z \right) = exp \left(-z \right),   \label{exp}
\end{equation}
which for $z=E/T$ (with $E$ being energy and the scale parameter $T$ being temperature) considered here is simply the Boltzmann-Gibbs (BG) distribution\footnote{The choice of the exponential distribution (\ref{exp}) as the final equilibrium distribution results from the following premises: it describes the spectra in the thermalized system, it is a memoryless probability distribution, it describes the processes in which events occur independently and it is the most probable distribution for the fixed expected value.}.

\begin{figure}[h]
\begin{center}
\includegraphics[scale=0.4]{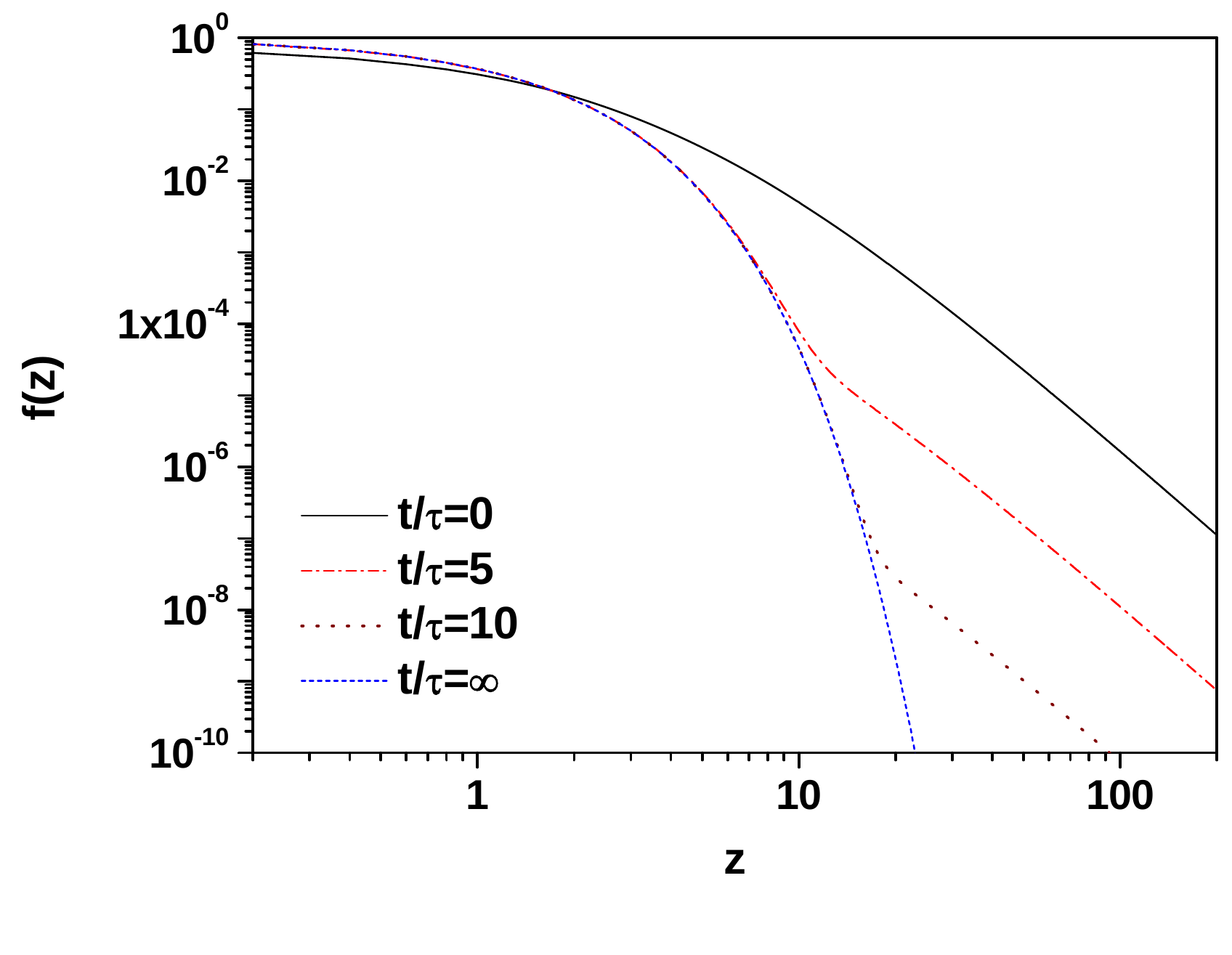}
\end{center}
\vspace{-8mm}
\caption{Schematic probability distributions $f_{fo}(z)$ resulting from the relaxation time approximation scenario calculated using Eq.(\ref{A-1}) for $f_{in}$ given by Eq.(\ref{power}) with $\gamma = 4$ and  $f_{eq}$  given by Eq.(\ref{exp}), for $t/\tau = 0$, $5$, $10$, $\infty$ (curves from top down).}
\label{FA1}
\end{figure}

Now note that the final distribution given by Eq. (\ref{RTA-sol}) can be rewritten in a two-component form,
\begin{equation}
f_{fo}(z) = f_{in}(z)\exp\left(-\frac{t_f}{\tau}\right) + f_{eq}(z)\left[ 1 - \exp\left(-\frac{t_f}{\tau}\right)\right]. \label{A-1}
\end{equation}
 This emphasizes that the sought two-component final distribution $f_{fo}$ is bistable and jumps between the BG distribution $f_{eq}$ and some assumed initial distribution $f_{in}$ with probability described by the parameter $t_f/\tau$. To illustrate this effect we use quasi-power like initial distribution $f_{in}(z)$
\begin{equation}
f_{in}\left( z\right) = C\cdot\left[ 1 + \frac{z}{\gamma}\right]^{-\gamma}, \label{power}
\end{equation}
with $\gamma=4$ and $C=(\gamma-1)/\gamma$ being the normalization factor, and the exponential distribution (\ref{exp}) for $f_{eq}$; for this choice we get $f_{fo}$  for different values of $t_f/\tau$ as presented in Fig. \ref{FA1}. The clearly visible structure of $f_{fo}(z)$ for some choice of parameters clearly illustrates the potential bi-stability of the final system being a consequence of its composition of quasi-power and exponential distributions.

Let us summarize this part. In the RTA  we always have at our disposal two distributions: the initial $f_{in}$ and  the exponential $f_{eq}$. In the case considered here at the beginning, i.e. for $t = 0$, all the objects of the system are described by the power distribution. Over time, some objects (decreasing with time) are still described by the power distribution, but the remainder (increasing with time) become thermalized and are described
by the exponential distribution. So: for a given $t$, every object of the system is either in the initial state or in the final state and the occupation numbers of these states changes with time.

\section{Beyond the RTA - time evolution of internal variable}
\label{sect.internal-variable}

Note however that a given non-equilibrium process, regardless of the specific dynamics, evolves the probability distribution over the system configurations. The system traverses a probability manifold, which is not a manifold of equilibrium states because the distribution at each point on the manifold need not correspond to a BG distribution \cite{NatPhys}. This means that for every $t$ one has some distribution $f(z,t)$ and all objects in the system are described by that (single) distribution. In our case this distribution should smoothly evolve with time from power to exponential (equilibrium) form, i.e., it should be a quasi-power like distribution. The most common distribution of this type is the Tsallis distribution \cite{T,T1,T3},
\begin{equation}
f(z,t) = (2-q)\left[ 1 + (q -1)z\right]^{\frac{1}{1-q}}, \label{TFit}
\end{equation}
characterized by the energy dependent parameter $q=q(t)$ (note that for $q\to 1$ the Tsallis distribution becomes the BG from Eq. (\ref{exp})). To justify such a choice let us recall a unique feature of distribution (\ref{TFit}) which distinguishes it from all other distributions used so far in this context. Namely, whereas for $z \to \infty$ (or for $z >> 1/(q-1)$) it becomes a power distribution, for $z \to 0$ (or for $z <<1/(q-1)$) it goes into an exponential distribution. Thus, for an appropriate selection of the parameter $q(t)$, it can describe all the distributions occurring in the RTA formula: $f_{in}(z,t)$ for $q\left(t_f=0\right) = q_{in} > 1$ and $f_{eq}$ for $q\left( t_f \to \infty\right) = 1$.

The effectiveness of the Tsallis distribution is best evidenced by the results of works \cite{WW1,WWCT,WW2}. In particular, as shown in \cite{WWCT}, it nicely describes a wide range of the measured transverse momenta ($0.1 < p_T < 100$ GeV which corresponds in this case to $1 < z < 700$) in which the cross section spans a range of $\sim 14$ orders of magnitude\footnote{Recently, inspired by the Tsallis statistics for a non-equilibrium distributions, the non-extensive BTE in the relaxation time approximation was discussed in \cite{qBTE}.}.

The time-dependent parameter $q = q (t)$ (more specifically, its deviation from $q=1$) represents the degree of the non-extensivity or, in other words, the degree of deviation of the system from the thermalized or equilibrated system, which is usually described by the well known BG statistical mechanics. It is also a control parameter that fully defines the shape of the Tsallis distribution, in particular its evolution over time through moments such as the expected value, $\langle z(t)\rangle$ and variance, $Var[z(t)]$:
\begin{eqnarray}
\langle z(t) \rangle &=& \frac{1}{3-2q(t)}, \label{mean}\\
Var(z) &=& \frac{2-q(t)}{(3-2q(t))^2(4-3q(t))}.    \label{var}
\end{eqnarray}
In general, the moments $<z^n>$, limited to $n+1<1/(q-1)$, are related by the recurrence relation
\begin {equation}
\langle z^n\rangle = \frac{n}{1-(n+1)(q-1)}\langle z^{n-1}\rangle.  \label{recurence}
\end{equation}

As mentioned above, we assume that the dynamic evolution of a system over time smoothly and monotonically transforms the probability distribution $f(z,t)$ in $(z,t)$-space and that $f(z,t)$ is a Tsallis distribution fully described by the time-dependent control parameter $q(t)$.

Now note that if we had completely formally used the Tsallis distributions for all distributions in
Eq. (\ref{RTA-sol}) defining the RTA and comparing $f(z=0)$ or $<z>$, we would get such a relationship between the parameters $q$ appearing there (remembering that $f_{eq}$ is assumed as a BG distribution which is equavalent to a Tsallis distribution with $q_{eq}=1$):
\begin{equation}
\left(2 - q_{fo}\right) = 1 + \left[ \left(2-q_{in}\right) - 1 \right]\cdot \exp \left(-\frac{t_f}{\tau}\right).
\label{sumrule}
\end{equation}

However, if from the very beginning we decide to describe the entire process using only quasi-power Tsallis distributions, the time evolution of which is given only by the time evolution of their control parameters $q=q(t)$, which means that $f (t) = f[q (t)]$, we should go back to Eq. (\ref{RBE}), which is now
\begin{equation}
\frac{\partial f(t)}{\partial t} = F[q(t)]. \label{A-2}
\end{equation}
This equation replaces  Eq. (\ref{RBE}). The form of the function $F$ from Eq. (\ref{A-2}) can be deduced by taking $f(t)$ given by the Tsallis distribution with $q = q(t)$ and calculating $df/dt$. As a result, we get that
\begin{eqnarray}
F\left[z,q(t)\right] &=& f\left[ z,q(t)\right]\cdot Q \cdot\frac{1}{[q(t)-1]^2}\frac{dq(t)}{dt}, \nonumber\\
Q &=& \ln\left[ 1 + (q(t)-1)z \right] + \frac{1}{1 + [q(t)-1]z} - \nonumber\\
&&-\frac{[q(t) - 1]^2}{2 - q(t)} - 1,\label{F}
\end{eqnarray}
and expressing the dependence of $Q(z)$ on the variable $z$ by the function $f=f[z,q(t),t]$ we have
\begin{eqnarray}
&&\frac{\partial f[z,q(t),t]}{\partial t} = -\frac{f}{\tau}\cdot \nonumber\\
&& ~~~~~\cdot \left\{ \frac{1}{q-1}\left[ \left( \frac{f}{2-q}\right)^{q-1} - \frac{1}{2-q}\right] - \ln\left( \frac{f}{2-q}\right)\right\}. \label{FF}
\end{eqnarray}

\begin{figure}[t]
\begin{center}
\includegraphics[scale=0.4]{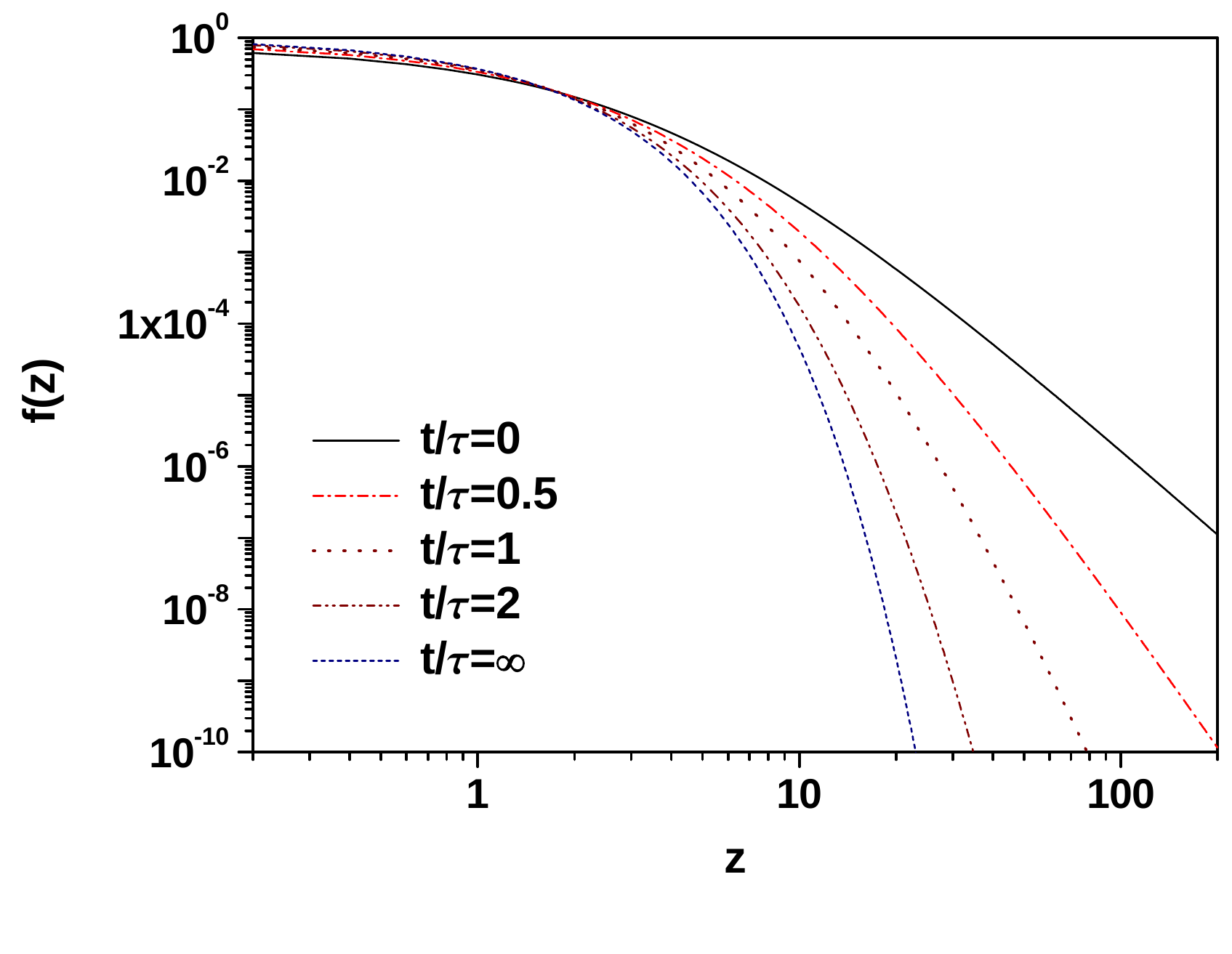}
\end{center}
\vspace{-8mm}
\caption{Schematic probability distributions  $f_ {fo}(z)$ calculated from Eq. (\ref{A-2}) for $t/\tau = 0$, $0.5$, $1$, $2$, $\infty$ (curves from top down). The distributions $f_{in}$ and $f_{eq}$ are the same as in Fig. \ref{FA1} (given by the Tsallis formula (\ref{TFit}) with $q_{in}=1.25$ and $q_{eq} = 1$, respectively).}
\label{FA2}
\end{figure}

To go further, we need to set the time dependence of the parameter $q$ in some way. Note that the nonextensivity parameter $q(t)$ describes deviations of the state of a statistical system from equilibrium and in this sense it plays the role of an {\it internal variable} discussed in Refs. \cite{IP,VP}. Therefore, following such an approach \cite{VP}  we assume that the equation of the dynamics describing the control parameter $q(t)$ is of the form of the equation of a relaxation:
\begin{equation}
\frac{d q}{d t} = \frac{q_{eq}-q}{\tau}. \label{qRTA}
\end{equation}
Remembering that we always assume that $q_{eq} =1$, the solution of (\ref{qRTA}) is
\begin{equation}
q - 1 = \left( q_{in} - 1\right) \exp\left( - \frac{t_f}{\tau}\right) \label{A-4}
\end{equation}
which coincides with Eq. (\ref{sumrule}). Fig. \ref{FA2} shows the resultant schematic distributions $f_{fo}$ for different $t_f/\tau$; they all have the form of a Tsallis distribution with $q=q\left(t=t_f\right)$ as given by
Eq. (\ref{A-4}). As one can see the result is now different from that using the RTA approximation shown in Fig. \ref{FA1}.

\begin{figure}[h]
\begin{center}
\includegraphics[scale=0.4]{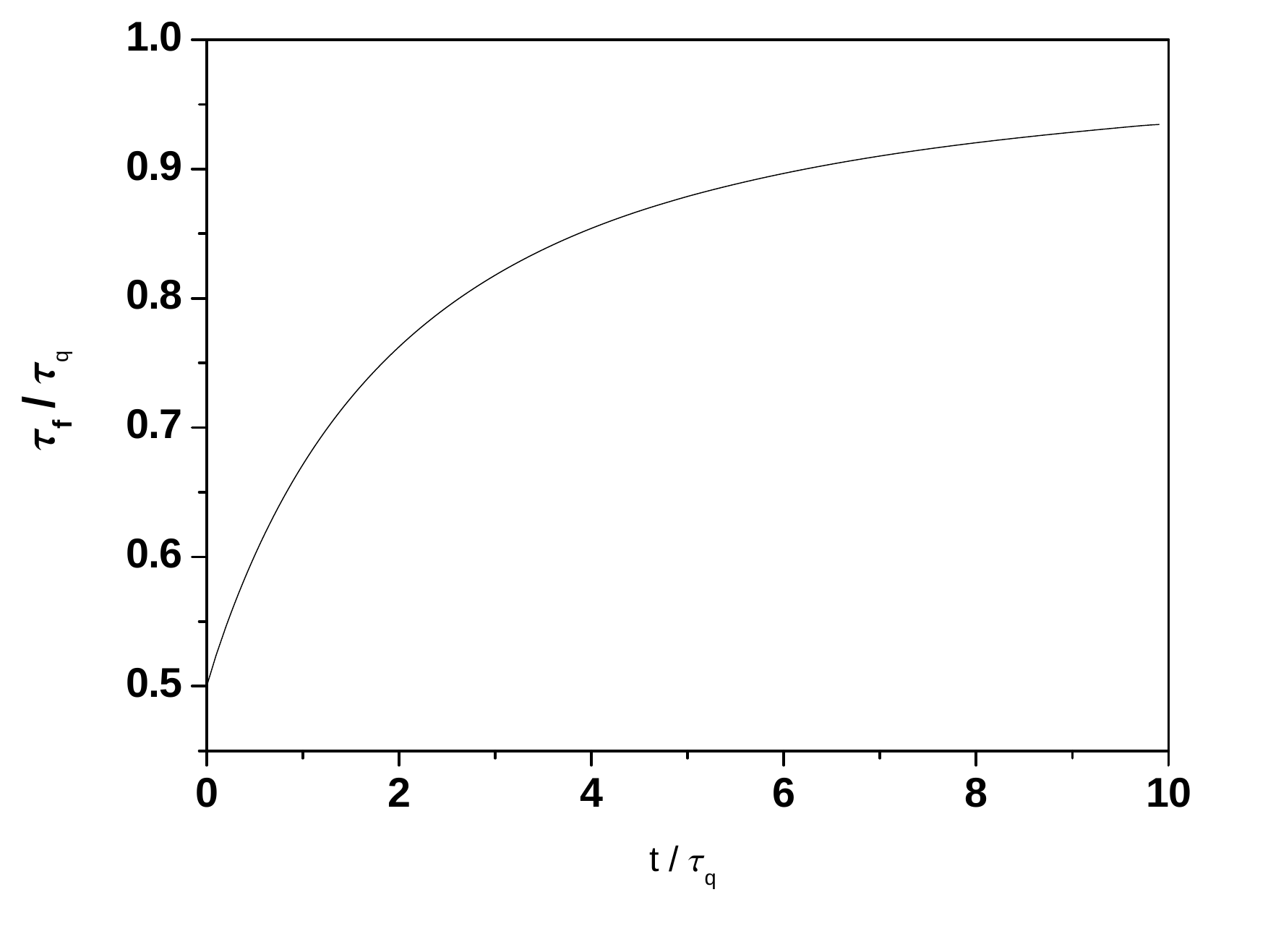}
\end{center}
\vspace{-8mm}
\caption{Time dependence of the the relaxation times ratio $ \tau_f / \tau_q $ for an initial mean value $ <z>_{in}=2$ (corresponding to $q_{in}=1.25$). }
\label{FA3}
\end{figure}

Notice that the relaxation times $\tau=\tau_f$ in Eq.(\ref{RTA}) and $\tau=\tau_q$ in Eq.(\ref{qRTA}) describe the relaxation of different quantities, respectively the entire distribution and its control parameter. Comparing (for the same time) mean values $\langle z\rangle = 1 + \exp\left( - \frac{t}{\tau_f}\right)\left( \langle z\rangle_{in} -1\right)$ evaluated from the RTA, Eq. (\ref{A-1}), and its $q$ version, $\langle z \rangle = \left[ 1 + \exp\left( -\frac{t}{\tau_q}\right)\left(\langle z\rangle_{in}^{-1} - 1 \right)\right]^{-1}$, Eq. (\ref{TFit}) for $q(t)$ given by Eq. (\ref{A-4}), we have that
\begin{equation}
\frac{t}{\tau_f}=\frac{t}{\tau_q}+\ln [<z>_{in}+(1-<z>_{in})\exp (-t/ \tau_q)] \label{tau_f_q}
\end{equation}
and the ratio of relaxation times in both approaches is
\begin{equation}
\frac{\tau_f}{\tau_q}=\frac{t/ \tau_q}{t/ \tau_q+ \ln [<z>_{in}+(1-<z>_{in}) \exp(-t/ \tau_q)]} \label{ratio}
\end{equation}
changing from $\tau_f / \tau_q = 1/<z>_{in}$ for $t\rightarrow 0$ to $\tau_f / \tau_q = 1$ for $t\rightarrow \infty$ (cf. Fig \ref{FA3}).

Note that in a situation where in some isolated system we have a fixed number of particles $N$ and a fixed total energy $U$, we have a constant average energy $\langle E \rangle$. Therefore, in such a case, the variability in time of $\langle z\rangle=\langle E\rangle/T$ must be caused by the appropriate variability in time of the scale parameter $T$ (here the temperature). This means that the scale parameter in our scaled variable $z=x/x_0$ also changes with time:
\begin {equation}
x_0(t) = x_0^* [1-2(q_{in}-1)\exp (-t/ \tau)], \label {mvt}
\end {equation}
where $x_0=x_0^*$ at equilibrium ($t\longrightarrow \infty$). Because, as known from \cite{WW},  fluctuations of the scale parameter $x_0$ are directly connected with the parameter $q$:
\begin{equation}
\frac{Var(1/x_0)}{<1/x_0>^2} = q - 1,  \label{q}
\end{equation}
we can write that
\begin{equation}
Var(1/x_0) = (q_{in}-1)\langle 1/x_0\rangle^2 \exp\left( - t/\tau\right). \label{Vrx0}
\end{equation}
This therefore means that the relaxation time $\tau$ now describes the temporal evolution of the fluctuations of the scale parameter $x_0$ (in the scaled variable $z=x/x_0$).

\section{Conclusions}
\label{sect.conclusion}

A few remarks that may inspire further research in this direction could be of interest here \cite{NatPhys}. Notice that in the language of information theory based on Shannon entropy,
\begin{equation}
S = - \int dz f(z) ln[f(z)], \label{SE}
\end{equation}
our  time evolution Eq. (\ref{A-2}) can be expressed as
\begin{equation}
\frac {\partial f}{\partial t}= - f \frac{\partial S_z}{\partial t}\label{evolution}
\end{equation}
where the $S_z = - \ln[f(z)]$ is  suprisal, which measures the information gained by observing the outcome $z$ in the system and the Shannon entropy is its expectation value. Now note that for the Tsallis distribution (\ref{TFit}) we have that $S_z \sim z$ and $\partial S_z/\partial t \sim z$. The entropy rate is given by
\begin {equation}
\frac {\partial S}{\partial t}= - \left< S_z \frac {\partial S_z}{\partial t} \right>. \label{rate}
\end{equation}
From \cite{NatPhys} we know that the linear relationship between $ \partial S_z / \partial t$ and $z$ guaranties that distribution (\ref{TFit}) saturates the time-information uncertainty bound:
\begin{equation}
\Bigg| Cov\left(\frac{\partial S}{\partial t},z\right)\Bigg| \leq \sqrt{Var\left( \frac{\partial S}{\partial t}\right)Var(z)}. \label{TINFOB}
\end{equation}
More precisely, for the Tsallis distribution
\begin{equation}
 Cov\left( \frac{\partial S_z}{\partial t}, z\right) = \left< \frac{\partial S_z}{\partial t} \cdot z\right> - \left< \frac{\partial S_z}{\partial t}\right> \langle z\rangle = 0 \label{COV}
\end{equation}
because $ \langle \partial S_z / \partial t \cdot z \rangle = 0$ and (for any distribution)  $\langle\partial S_z / \partial t\rangle =0$.
The distance of a given distribution (in our case defined by $q(t)$) from the equilibrium distribution (defined by $q=1$) is given by the difference of the corresponding entropies:
\begin {eqnarray}
S[q(t)]\!-\! S(q=1)\!\!\! &=&\!\!\! \frac{q(t)\!-1\!}{1\!-\![q(t)-1)]}-\ln \{1\!-\! [q(t)\!-\! 1]\} \simeq \nonumber \\
&\simeq& \frac{2[q(t) - 1]}{2 - q(t)}.  \label{d-S}
\end{eqnarray}
From equations (\ref{FF}), (\ref{qRTA}) and (\ref{d-S}) we get that
\begin{equation}
\frac{d f(t)}{dt} = \frac{f(t)}{\tau}\cdot\frac{q(t) - 1}{2 - q(t)} = \frac{f(t)}{\tau}\cdot \frac{S[q(t)] - S(q=1)]}{2}. \label{Fin}
\end{equation}

In conclusion, we propose a new, modified form for the relaxation time approximation for the collision term in the Boltzmann equation (\ref{RBE}) allowing a smooth transition to the thermalized distribution. It consists of replacing the simple form of this term, given by the Eq. (\ref{RTA}), where the relaxation time $\tau$ determines how fast the equilibrium state of the studied distribution $f$ is reached, by Eq. (\ref{qRTA}) describing the time evolution of the most important (control) parameter of the analyzed distribution. The relaxation time $\tau$ would now control the rate of change of this parameter from some value to one that corresponds to the equilibrium state. We argue that this is possible if we use for the phenomenological description of the distributions of interest the quasi-power law Tsallis distribution given by Eq. (\ref{TFit}) which is able to describe a given process at all stages of time evolution. Its control parameter is the time-dependent non-extensivity parameter $q(t)$ and the relaxation time parameter $\tau$ describes its time evolution,  as shown in Eq. (\ref{qRTA}).  This  single quasi-power law probability distribution (\ref{TFit}) smoothly evolving towards thermalization would then replace the two-component distribution given by Eq. (\ref{A-1}) which arises from the RTA. The proposed scheme offers multiple applications in situations where one wants to study the time evolution of an ensemble but one does not want to invoke the kinetic theory with complicated collision integrals.

\begin{acknowledgements}
This research was supported by the Polish National Science Centre grant UMO-2016/22/M/ST2/00176 and by the Polish Ministry of Science and Higher Education grant DIR/WK/2016/2018/17-1 (GW). We thank Dr Nicholas Keeley for reading the manuscript.
\end{acknowledgements}


\end{document}